\documentclass[twocolumn,showpacs,amsmath,amssymb,floatfix]{revtex4}

\usepackage{graphicx}
\usepackage{color}

\begin{document}

\title{Diagrammatic determinantal quantum Monte Carlo
           methods:\\  Projective schemes and applications to
           the Hubbard-Holstein model}
\author{F.~F.~Assaad and T.~C.~Lang}
\affiliation{Institut f\"ur theoretische Physik und Astrophysik,
Universit\"at W\"urzburg, Am Hubland D-97074 W\"urzburg}

\date{\today}

\begin{abstract}
We extend the weak-coupling diagrammatic determinantal algorithm to projective schemes
as well as to the inclusion of phonon degrees of freedom. The projective approach
provides a very efficient algorithm to access zero temperature properties. To
implement  phonons, we integrate them out in favor of a retarded
density-density interaction and simulate the resulting purely electronic action with
the weak-coupling diagrammatic determinantal algorithm. Both extensions are tested
within the dynamical mean field approximation for the Hubbard and Hubbard-Holstein
models.
\end{abstract}


\maketitle

\section{Introduction}

Diagrammatic determinantal quantum Monte Carlo (DDQMC), be it the weak-coupling expansion
\cite{Rubtsov05}, or hybridization expansion \cite{Werner06} approach, is emerging as the
method of choice for impurity solvers \cite{Gull06}. In comparison to the Hirsch-Fye
approach  \cite{HirschFye86} they are continuous time methods and thereby free of
Trotter errors, more efficient, and more flexible. In this article we concentrate on the
weak-coupling algorithm. After a short review of our implementation of the algorithm,
we  show how to generalize it to projective schemes as well as to the inclusion of
phonon degrees of freedom. 

Projective schemes have already been implemented in the framework of the Hirsch-Fye
algorithm and used in the context of dynamical mean field theories \cite{Feldbach04,Feldbach04_com,Feldbach04_rep}. 
Very similar ideas for the formulation of a projective DDQMC algorithm may be used and
are reviewed in Sec. \ref{Proj_DDQMC}. With the projective DDQMC, we can reproduce
results of \cite{Feldbach04} at a fraction of the computational cost and access much
lower projection parameters. 

Phonon degrees of freedom have very recently been implemented in the hybridization
formulation of the DDQMC \cite{Werner07}. Since the hybridization approach is based on
the expansion in the hybridization, the inclusion of phonons relies on a Lang-Firsov
transformation. In the weak coupling approach it is more convenient to integrate out the
phonons in favor of a retarded interaction. The purely electronic model may then be
solved efficiently within the weak coupling DDQMC. In Sec. \ref{Phonons_DDQMC} we
present some details of the algorithm and provide test simulations for the Hubbard-Holstein 
model in the dynamical mean-field theory (DMFT) approximation.\\

\section{The diagrammatic determinantal method for Hubbard interactions}\label{Diag_basic}

Here we will briefly review the diagrammatic determinantal method for the Hubbard model
\begin{equation}
   \hat{H}_U =
      \underbrace{-\sum_{ i,j,\sigma} t_{i,j} \hat{c}^{\dagger}_{i,\sigma} 
                                              \hat{c}_{j,\sigma}}_{\hat{H}_0}  
    +             U \sum_{i} \left(\hat{n}_{i,\uparrow}   -1/2 \right)
                             \left(\hat{n}_{i,\downarrow} -1/2 \right) \;,
\end{equation}
where $\hat{n}_{i,\sigma} = \hat{c}^{\dagger}_{i,\sigma}\hat{c}_{i,\sigma} $ and 
$\hat{c}^{\dagger}_{i,\sigma}$ ($\hat{c}_{i,\sigma}$) creates (annihilates) a fermion in
a Wannier state centered around site $i$ ($j$) and with $z$-component of spin $\sigma$.

As will become apparent in subsequent sections, we rewrite the Hubbard interaction as 
\begin{eqnarray}
   \label{Ising}
	\frac{U}{2} \sum_{i} \sum_{s = \pm 1}  
      \left(\hat{n}_{i,\uparrow} -1/2 - s \delta \right) 
      \left(\hat{n}_{i,\downarrow} -1/2 + s \delta \right) \;.
\end{eqnarray} 
to avoid the negative sign problem at least for impurity and one-dimensional models.
After carrying out the sum over the Ising spins, $s$, one recovers the original
Hubbard interaction  up to a constant. As will be seen below an adequate choice of
$\delta$ to avoid the sign problem for a one-dimensional chain reads ${\delta =
\frac{1}{2} + 0^+}$.

A weak coupling perturbation expansion yields for the partition function: 
\begin{widetext}
   \begin{equation}
      \frac{Z}{Z_0} = \sum_{n=0}^{\infty} \left(\frac{-U}{2}\right)^n 
         \int_{0}^{\beta} {\rm d} \tau_1 \sum_{i_1,s_1} \cdots
         \int_{0}^{\tau_{n-1}} {\rm d} \tau_n  \sum_{i_n,s_n}
         \prod_{\sigma} \langle
            \left[\hat{n}_{i_1,\sigma}(\tau_1) -\alpha_{\sigma}(s_1)\right]\cdots 
            \left[\hat{n}_{i_n,\sigma}(\tau_n) -\alpha_{\sigma}(s_n)\right]
         \rangle_0 \;.
   \end{equation}
\end{widetext}
Here, we have defined
\begin{equation}
   \alpha_{\sigma} (s) = 1/2 + \sigma s \delta 
\end{equation}
and $\langle\bullet\rangle_0 = {\rm Tr\,}\left[e^{-\beta \hat{H}_0}\bullet
\right]/Z_0$ with $Z_0={\rm Tr\,} \left[e^{-\beta\hat{H}_0}\right]$. Note that the Ising
field $s$ has obtained an additional time index. The thermal expectation value is the
sum over all diagrams, connected and disconnected, of a given order $n$. Using Wick's
theorem this sum can be expressed as a determinant where the entries are the Green's
functions of the non-interacting system. 
\begin{widetext}
   \begin{eqnarray}
      \label{Det}
      \langle T\left[\hat{n}_{\sigma,i_1}(\tau_1)-\alpha_{\sigma}(s_1)\right]\cdots 
               \left[\hat{n}_{\sigma,i_n}(\tau_n)-\alpha_{\sigma}(s_n)\right]
      \rangle_0 = \qquad\qquad\qquad\qquad\qquad\qquad\qquad\qquad\:\nonumber\\
      \det\left( 
         \begin{array}{cccc}
            G^0_{i_1,i_1}(\tau_1,\tau_1)-\alpha_{\sigma}(s_1) & G^0_{i_1,i_2}(\tau_1,\tau_2)                                 & \cdots & G^0_{i_1,i_n}(\tau_1,\tau_n)                                \\
            G^0_{i_2,i_1}(\tau_2,\tau_1)                                 & G^0_{i_2,i_2}(\tau_2,\tau_2)-\alpha_{\sigma}(s_2) & \cdots & G^0_{i_2,i_n}(\tau_2,\tau_n)                                \\ 
           \cdot                                                         & \cdot                                                        & \cdot  & \cdot                                                       \\
           \cdot                                                         & \cdot                                                        & \cdot  & \cdot                                                       \\
           \cdot                                                         & \cdot                                                        & \cdot  & \cdot                                                       \\
           G^0_{i_n,i_1}(\tau_n,\tau_1)                                  & G^0_{i_n,i_2}(\tau_n,\tau_2)                                 & \cdots & G^0_{i_n,i_n}(\tau_n,\tau_n)-\alpha_{\sigma}(s_n)
         \end{array} 
      \right)\;,
   \end{eqnarray}
\end{widetext}
with Green's functions
\begin{equation}
   G^0_{i,j}(\tau_1,\tau_2) = \langle T \hat{c}^{\dagger}_i(\tau_1)\hat{c}_j(\tau_2) \rangle_0 \;,
\end{equation}
which we have assumed to be spin independent. In the above, $T$ corresponds to the time
ordering. Defining a configuration, $C_n$, by the $n$ Hubbard vertices, as well as the
Ising spins introduced in Eq.~(\ref{Ising})
\begin{equation}
\label{Conf}
	C_n = \left\{\left[i_1,\tau_1,s_1\right]\cdots 
                \left[i_n,\tau_n,s_n\right]\right\} \;,
\end{equation}
and the {\it sum} over the configuration space by
\begin{equation}
   \label{Sum_Conf}
	\sum_{C_n} = \sum_{n=0}^{\infty}  
      \int_{0}^{\beta}{\rm d}\tau_1\sum_{i_1,s_1}\cdots
      \int_{0}^{\tau_{n-1}}{\rm d}\tau_n\sum_{i_n,s_n} \;,
\end{equation}
the partition function can conveniently be written as
\begin{equation} 
\label{Weight}
	\frac{Z}{Z_0} = \sum_{C_n}  \left(-\frac{U}{2} \right)^n \prod_{\sigma} \det M_{\sigma}(C_n)\;.
\end{equation}
Here $M_{\sigma}$ is the $n\times n$ matrix of Eq.~(\ref{Det}). Observables,
$\hat{O}(\tau)$, can now be computed with
\begin{equation}
   \langle\hat{O}(\tau)\rangle =
      \frac{\sum_{C_n}\left(-\frac{U}{2}\right)^n\prod_{\sigma}\det M_{\sigma}(C_n)\langle\langle\hat{O}(\tau)\rangle\rangle_{C_n}}
           {\sum_{C_n}\left(-\frac{U}{2}\right)^n\prod_{\sigma}\det M_{\sigma}(C_n)}\;,
\end{equation}
where for $\hat{O}(\tau)=\prod_{\sigma}\hat{O}_{\sigma}(\tau)$ we have
\begin{widetext}
   \begin{equation}
   \label{Obs}
	\langle\langle \hat{O}(\tau)\rangle \rangle_{C_n} = 
      \frac{\prod_{\sigma}\langle T
         \left[\hat{n}_{i_1,\sigma}(\tau_1)-\alpha_{\sigma}(s_1)\right]\cdots 
         \left[\hat{n}_{i_n,\sigma}(\tau_n)-\alpha_{\sigma}(s_n)\right]\hat{O}_{\sigma}(\tau)\rangle_0}
           {\prod_{\sigma}\langle T
         \left[\hat{n}_{i_1,\sigma}(\tau_1)-\alpha_{\sigma}(s_1)\right]\cdots 
         \left[\hat{n}_{i_n,\sigma}(\tau_n)-\alpha_{\sigma}(s_n)\right]\rangle_0} \;.
   \end{equation}
\end{widetext}
For any given configuration of vertices $C_n$, Wick's theorem holds. Hence, any
observable can be  computed from the knowledge of the single particle Green's function
\begin{eqnarray}
   \langle\langle T c^{\dagger}_{i,\sigma}(\tau)c_{j,\sigma}(\tau_1)
   \rangle\rangle_{C_n}
      = G^{0}_{i,j}(\tau,\tau_1)\qquad\qquad\qquad\nonumber\\  
      - \sum_{r,s=1}^{n} G^{0}_{i,i_r}(\tau,\tau_{i_r})
        \left(M_{\sigma}^{-1}\right)_{r,s} G^{0}_{i_s,j}(\tau_s,\tau_1) \;.
\end{eqnarray}
Here, we have assumed that the non-interacting Green's functions are spin independent.
As a consequence of the above equation, it becomes apparent that one can measure
directly the Matsubara Green's functions. This aspect facilitates the implementation of
the algorithm within the framework of dynamical mean-field theories.

\subsection{Sign problem}\label{Sign}

In auxiliary field determinantal methods \cite{Assaad02} it is known that the presence of
particle-hole symmetry can be used to avoid the negative sign problem. An identical
statement holds for the diagrammatic determinantal method.  We assume that $\hat{H}_0$ 
is invariant under the particle-hole transformation
\begin{equation}
    c^{\dagger}_{i,\sigma} \rightarrow (-1)^{i} c_{i,\sigma}. 
\end{equation}
As a consequence, 
\begin{eqnarray}
  \langle T  \prod_{r=1}{n} \left[\hat{n}_{i_r,\sigma}(\tau_r)-\alpha_{\sigma}(s_r)  \right]  
        \rangle_0 = 
    \qquad\qquad  \\
    (-1)^n
  \langle T \prod_{r=1}{n} \left[\hat{n}_{i_r,\sigma}(\tau_r)-\alpha_{-\sigma}(s_r)\right] \rangle_0 \;\;\; 
\nonumber
\end{eqnarray}
such that   ${\det M_{\uparrow} = (-1)^n \det M_{\downarrow}}$. A 
glimpse at Eq.~(\ref{Weight}) will confirm the absence of sign problem for this special case.
The above results is independent on the choice of $\delta$ introduced in
Eq.~(\ref{Ising}).  As we will see  in Sec. \ref{Tests} 
the algorithm is optimal at $\delta =0$. In this 
special case, $\det M_{\uparrow} = \det M_{\downarrow} = (-1)^n \det M_{\downarrow}$ such that 
only even values of $n$ occur in the sampling. We note that this vanishing of the weight for odd 
values of $n$ can be avoided by choosing a small value of $\delta$.

In one dimension and in the absence of frustrating interactions, there is no negative
sign problem \footnote{ To be more precise, configurations with negative weights do
occur. However, those configurations stem from the real space winding of the fermions
and can be eliminated  if open boundary conditions are adopted.}. The diagrammatic
approach also satisfies this property, provided that we choose  ${\delta = 1/2 + 0^+}$.
The quantity $\prod_{\sigma}\det M_{\sigma}(C_n)$ in Eq.~(\ref{Weight}) is nothing but 
\begin{eqnarray}
\label{WL}
   {\rm Tr\,}
       \Big[e^{-\beta \hat{H}_0} \prod_{\sigma}
       \left[\hat{n}_{i_1,\sigma}(\tau_1) - \alpha_{\sigma}(s_1)\right]\Big.\quad\qquad\qquad\nonumber\\
       \cdots \Big.\left[\hat{n}_{i_n,\sigma}(\tau_n) - \alpha_{\sigma}(s_n)\right]\Big]
       \Big/{\rm Tr\,}\Big[e^{-\beta \hat{H}_0}\Big] \;,
\end{eqnarray}
which we can compute within the real-space world-line approach 
\cite{Hirsch81,Assaad91}. Here, each world line configuration has a positive weight.
Let us consider an arbitrary world-line configuration, and a site $(i,\tau)$  in the
space-time lattice. Irrespective if this site is empty, singly or doubly occupied the 
expectation value of the operator ${\prod_{\sigma}\left[ \hat{n}_{i,\sigma}(\tau)
-\alpha_{\sigma}(s)\right]}$ will take a negative  value. Recall that we have
set ${\delta = 1/2 + 0^+}$. Hence, for each world line configuration the expectation value of 
the operator
${\prod_{\sigma}\left[\left(\hat{n}_{i,\sigma}(\tau)-\alpha_{\sigma}(s_1)\right)\cdots
\left(\hat{n}_{i_n,\sigma}(\tau_n)-\alpha_{\sigma}(s_n)\right)\right]}$ has a
sign equal to $(-1)^n$. Summation over all world line configurations yields the
expression in Eq.~(\ref{WL}) which in turn has a sign $(-1)^n$. This cancels the sign of
the factor $ (-U/2)^n $ in Eq.~(\ref{Weight}), thus yielding an overall positive
weight. 

In the rewriting of the Hubbard term (see Eq.~(\ref{Ising})) we have introduced a new
dynamical Ising field so as to avoid the negative sign problem at least for the
one-dimensional Hubbard model. Alternatively, one can choose a static Ising field and 
compensate for it by a redefinition of $\hat{H}_0$. Such a static procedure is
introduced in \cite{Rubtsov05}.  For the class of models  considered,  we 
have not noticed substantial  differences in performance  between static and dynamical choices  of 
Ising fields. We however favor the dynamical version since it allows one to keep the 
SU(2) spin invariant form of the non-interacting Hamiltonian $\hat{H}_0$.  

\subsection{Monte Carlo Sampling} \label{Monte-Carlo}

In principle two moves, the addition and removal of Hubbard vertices, are sufficient 
\footnote{Clearly, those move are not sufficient for the particle-hole symmetric case and $\delta =0$. 
In this case and as argued in Sec. \ref{Sign},  the weights vanish  for odd values of $n$, and 
hence, the algorithm would not be ergodic. One can circumvent this problem by  (i) introducing moves 
which add or remove pairs of  vertices  or (ii) using a small value of $\delta$. For the simulations 
presented here we have opted for solution (ii) and chosen in general $\delta = 0.1$ at the expense 
 of a small performance loss. }. 
In the Metropolis scheme, the acceptance ratio for a given move reads 
\begin{equation} 
	P_{C\rightarrow C'} = {\rm min}\left(\frac{T^{0}_{C'\rightarrow C } W(C')}
                                             {T^{0}_{C \rightarrow C'} W(C) },1\right)\;. 
\end{equation}
where ${T^{0}_{C'\rightarrow C}}$ corresponds to the probability of proposing a move
from configuration $C'$ to configuration $C$ and $W(C)$ corresponds to the weight of the
configuration. To add a vertex  ${T^{0}_{C_n\rightarrow C_{n+1}} = \frac{1}{2 N \beta}}$
which corresponds to the fact that one has to pick at random an imaginary time in the
range $[0,\beta]$, a site $i$ in the range ${1 \ldots N}$ (with $N$ the number of sites)
as well as an Ising spin. The proposal probability to remove a vertex ${T^{0}_{C_{n+1}
\rightarrow C_{n} } = \frac{1}{n+1}}$ corresponds to the fact that one will choose at
random one of the $n+1$ vertices present in configuration $C_{n+1}$, hence
\begin{eqnarray}
   P_{C_{n} \rightarrow C_{n+1} } & = & {\rm min} \left(-\frac{U \beta N}{(n+1)} 
      \frac{\prod_{\sigma} \det M_{\sigma}(C_{n+1})}
           {\prod_{\sigma} \det M_{\sigma}(C_n    )},1\right) \nonumber\\
   P_{C_{n+1} \rightarrow C_{n} } & = & {\rm min} \left(-\frac{(n+1)}{U \beta N} 
      \frac{\prod_{\sigma} \det M_{\sigma}(C_n    )}
           {\prod_{\sigma} \det M_{\sigma}(C_{n+1})},1\right) \;.\nonumber\\
\end{eqnarray}
Apart from the above addition and removal of vertices, we have implemented moves which 
flip the Ising spins at constant order $n$ as well as updates which move Hubbard
vertices both in space and time.

\subsection{Tests}  \label{Tests}

\begin{figure}[t]
   \includegraphics[width=0.4\textwidth]{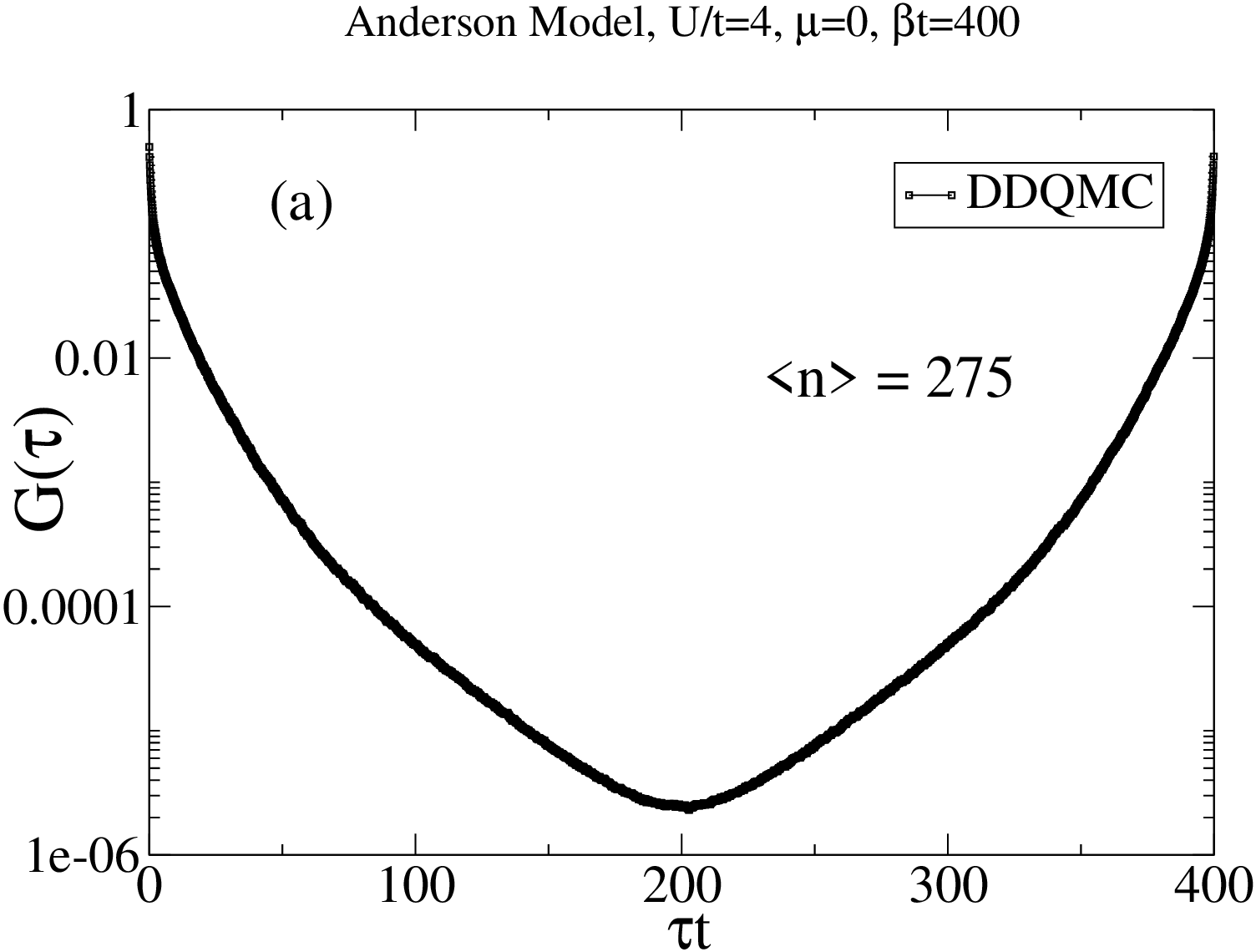} \\
   \includegraphics[width=0.4\textwidth]{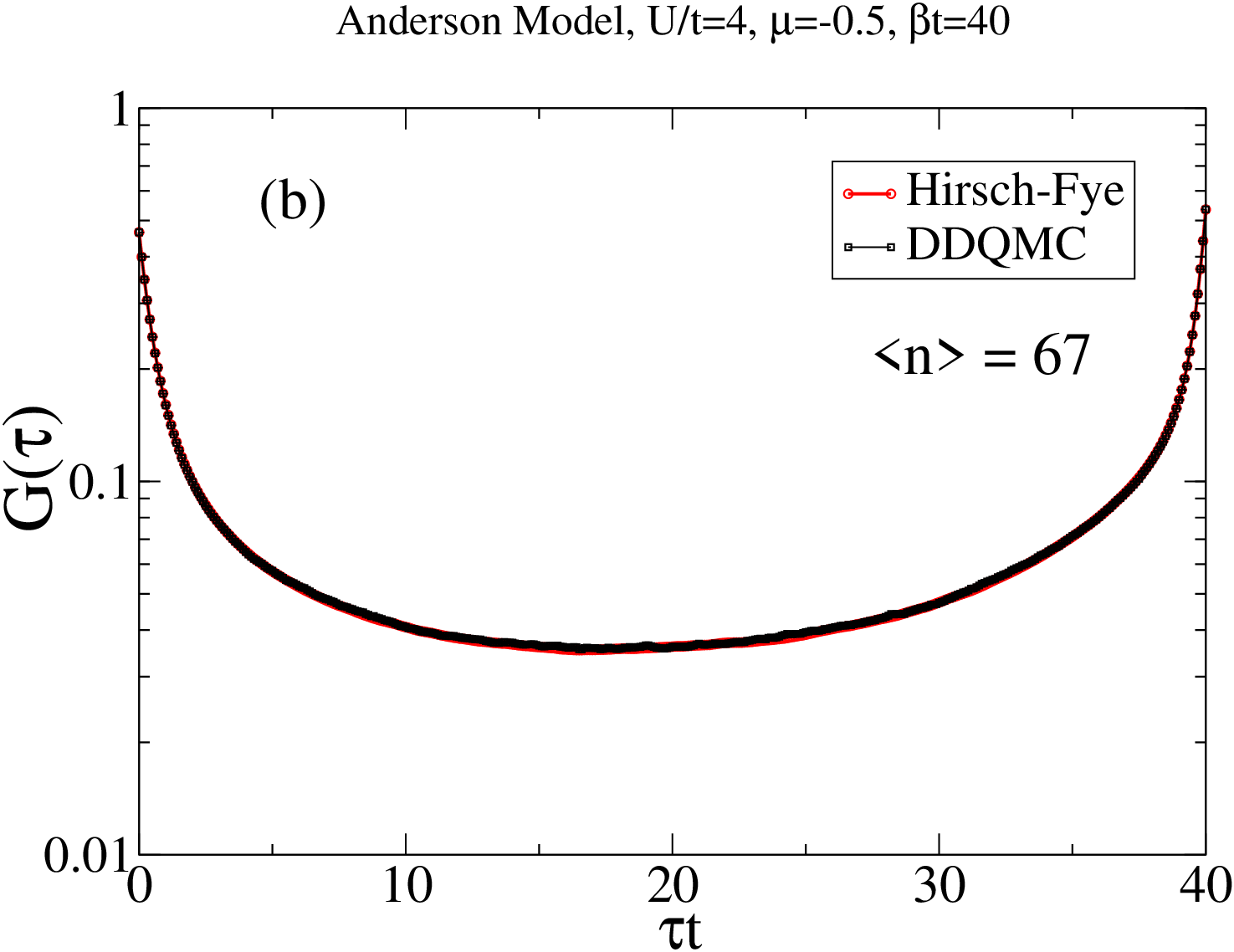} \\
   \caption{(Color online) Green's function $G(\tau)$ for the Anderson model of Eq.~(\ref{Siam}).
            (a) Particle-hole symmetric point, $\mu = 0$. To avoid the vanishing of the weight 
            at odd values of $n$ we have used $\delta = 0.1$ for this simulation. 
            (b) Away from particle-hole
            symmetry  $\mu = -0.5$.  For this higher temperature we provide a comparison
            with the Hirsch-Fye algorithm with Trotter step $\Delta \tau t = 0.1 $. 
            Note that in both cases we have used a $32 \times 32$ square lattice to
            generate the  non-interacting Green's function $G^{0}(\tau)$. $\langle n
            \rangle$ corresponds to the  average order expansion parameter.}
\label{Siam.fig}
\end{figure}
The efficiency of the approach relies on the autocorrelation time, which has to be
analyzed  on a case to case basis,  as well as on the  average expansion order
parameter. For a general interaction term $\hat{H}_1$ the average expansion parameter
is given by
\begin{eqnarray}
   \langle n \rangle
      & = & \frac{1}{Z} \sum_n \frac{(-1)^n n}{n!}
               \int_{0}^{\beta} {\rm d}\tau_1 \cdots \int_{0}^{\beta} {\rm d}\tau_n \nonumber\\
      &   &    \times\; \langle T \hat{H}_1(\tau_1) \cdots \hat{H}_1(\tau_n) \rangle_0 \nonumber\\ 
      & = & \frac{-1}{Z} \sum_m \frac{(-1)^m }{m!}
               \int_{0}^{\beta} {\rm d}\tau_1 \cdots \int_{0}^{\beta} {\rm d}\tau_m \int_{0}^{\beta} {\rm d}\tau \nonumber\\ 
      &   &    \times\; \langle T \hat{H}_1(\tau_1) \cdots \hat{H}_1(\tau_m) \hat{H}_1(\tau) \rangle_0 \nonumber \\
      & = &  - \int_{0}^{\beta}{\rm d} \tau \langle \hat{H}_1( \tau ) \rangle \;.
\end{eqnarray}
For the Hubbard model and replacing $\hat{H}_1$ by the form of Eq.~(\ref{Ising}) we
obtain:
\begin{equation}
   \label{Ave_n}
   \langle n \rangle = - \beta U \sum_i
      \left[\langle (\hat{n}_{i,\uparrow} -1/2)(\hat{n}_{i,\downarrow} -1/2)\rangle
          - \delta^2 \right] \;.
\end{equation}
Using the same techniques as in auxiliary field QMC methods \cite{Assaad02} the CPU costs for
the calculation of the acceptance probability and the update for the addition or removal of 
a vertex scales
as $n^2$. As apparent from Eq.~(\ref{Ave_n}) a sweep consisting of updating all $n$ vertices
results in an effort of $n^3$. Even though in this method ${M^{-1}_{\sigma}}$
is far better conditioned than in the \textit{classic} determinantal methods
\cite{HirschFye86,Blankenbecler81,Assaad02}, it has to be recalculated from scratch after several
updates which involves an effort of the order $n^3$ 
\footnote{  For the impurity problems  presented here, ${M^{-1}_{\sigma}}$ remains  very stable, such 
that recalculation from scratch of ${M^{-1}_{\sigma}}$ is not an issue.  
In contrast when applying the method to a lattice problem, we find that round-off errors 
cumulate and a more frequent recalculation of ${M^{-1}_{\sigma}}$  is required.}. 
Hence, because of these two limiting factors
the CPU time scales as ${(\beta\, U N)^3}$ which is precisely the same scaling as in the Hirsch-Fye
approach. Apart from the absence of Trotter errors the advantage of the method lies in a
large pre-factor. In the very special case of a single impurity, $N=1$, and for a
particle-hole symmetric Hamiltonian $\hat{H}_0$, the speedup is dramatic. Particle-hole
symmetry allows one to set ${\delta = 0}$ such that ${\langle n \rangle < \beta\,U/4}$. To
obtain this upper bound we have set the double occupancy to zero. Hence a simulation at ${U/t
= 4}$ and ${\beta\, t = 400}$ has a maximal average order parameter ${\langle n \rangle = 400
}$. In a Hirsch-Fye approach, one could opt for a Trotter step $\Delta \tau t = 1/8 $ and
hence $3200$ Trotter slices which determines the size of the matrices involved in the
simulations. Hence an underestimate of the speedup reads ${(3200 /400 )^3 = 512}$. Away from
particle-hole symmetry, the speedup is less impressive since we have to set ${\delta = 1/2 +
0^{+}}$ to avoid the negative sign problem. We have confirmed the above statements for the
Anderson impurity model,
\begin{eqnarray}
\label{Siam}
	\hat{H} & = & \sum_{k,\sigma} ( \epsilon(k) - \mu) 
                 \hat{c}^{\dagger}_{k,\sigma} \hat{c}_{k,\sigma}
                +\frac{V}{\sqrt{N}} \sum_{k,\sigma} \left( 
                 \hat{c}^{\dagger}_{k,\sigma} \hat{d}_{\sigma} + H.c.\right)\nonumber \\
           &  & + U \left( \hat{d}_{\uparrow}^{\dagger} \hat{d}_{\uparrow} -1/2 \right)
    \left( \hat{d}_{\downarrow}^{\dagger} \hat{d}_{\downarrow} -1/2 \right),
\end{eqnarray}
with Hubbard interaction $U$ and hybridization $V$. Here $\hat{c}^{\dagger}_{k,\sigma}$
($\hat{c}_{k,\sigma}$) creates (annihilates) a fermion in the conduction band at
momentum $k$ with spin $\sigma$ and the dispersion relation ${\epsilon(k) =-2t(\cos(k_x)
+ \cos(k_y))}$. The operators $\hat{d}^{\dagger}_{\sigma}$ and $\hat{d}_{\sigma}$
create, annihilates an impurity electron, respectively. Our results including comparison
with the Hirsch-Fye algorithm are presented in Fig.~\ref{Siam.fig}.

Finally let us note that applying the method to a one-dimensional Hubbard model of length
$N$, yields a very poor performance in comparison to standard finite temperature BSS  
auxiliary field algorithms \cite{Blankenbecler81} since those methods scale as 
${\beta \,U N^{3}}$ \cite{Assaad02}.

\section{Generalization to projective Approaches}\label{Proj_DDQMC}

Projective approaches rely on the filtering out of  the ground state, ${|\Psi_0\rangle}$ 
from a trial wave function $| \Psi_T \rangle$, which is required to be non-orthogonal to 
${|\Psi_0\rangle}$
\begin{eqnarray}
	\frac{\langle\Psi_0 | \hat{O} | \Psi_0\rangle }{\langle\Psi_0 | \Psi_0\rangle} 
   = \lim_{\Theta \rightarrow \infty} 
   \frac{\langle\Psi_T | e^{-\frac{\Theta}{2} \hat{H} } \hat{O} 
                         e^{-\frac{\Theta}{2} \hat{H} } |\Psi_T\rangle}
        {\langle\Psi_T | e^{-\Theta\hat{H}} | \Psi_T\rangle}\;.\nonumber\\
\end{eqnarray}
For convenience and simplicity, we will assume that  ${|\Psi_T\rangle}$ is the ground state
of $\hat{H}_0$, such  that for a given value of $\Theta$ the right hand side of the  above
equation can be written as
\begin{eqnarray}
   \frac{\langle \Psi_T | e^{-\frac{\Theta}{2} \hat{H} } \hat{O} 
                          e^{-\frac{\Theta}{2} \hat{H} } | \Psi_T \rangle }
        {\langle \Psi_T | e^{- \Theta \hat{H} }  | \Psi_T \rangle } \qquad\qquad\qquad\qquad\qquad\nonumber\\
  	 = \lim_{\beta_0 \rightarrow \infty} 
        \frac{ {\rm Tr\,} e^{-\beta_0 \hat{H}_0} 
                          e^{-\frac{\Theta}{2}\hat{H} } \hat{O}
                          e^{-\frac{\Theta}{2} \hat{H} } }
              {{\rm Tr\,} e^{-\beta_0 \hat{H}_0} e^{- \Theta \hat{H} } }\;.
\end{eqnarray}
With the definition ${Z_{p}={\rm Tr\,}\left[ e^{-\beta_0\hat{H}_0} e^{- \Theta\hat{H}}\right]}$
and ${Z_{p,0}={\rm Tr\,}\left[ e^{-(\beta_0+\Theta )\hat{H}_0}\right]}$ a weak
coupling expansion yields
\begin{widetext} 
   \begin{equation}
      \label{proj}
	   \frac { Z_{p} } {Z_{p,0} }  = 
         \sum_{n=0}^{\infty} \left(\frac{-U}{2}\right)^n 
         \int_{0}^{\Theta} {\rm d} \tau_1 \sum_{i_1,s_1} \cdots
         \int_{0}^{\tau_{n-1}} {\rm d} \tau_n  \sum_{i_n,s_n} \prod_{\sigma}
         \langle\left[
            \hat{n}_{i_1,\sigma}(\tau_1)-\alpha_{\sigma}(s_1)\right] \cdots 
      \left[\hat{n}_{i_n,\sigma}(\tau_n)-\alpha_{\sigma}(s_n)\right] \rangle_{p,0} \;,
   \end{equation}
\end{widetext}
where ${\langle\bullet\rangle_{p,0}={\rm
Tr\,}\left[e^{-(\beta_0+\Theta)\hat{H}_0}\bullet\right]/Z_{p,0}}$. The similarity to the
finite temperature algorithm is now apparent. We use Wick's theorem to express the
expectation value on the right hand side of Eq.~(\ref{proj}) in terms of the product of
two determinants. Taking the limit ${\beta_0 \rightarrow \infty}$ we obtain 
\begin{eqnarray}
   \langle\Psi_T | e^{- \Theta \hat{H} } | \Psi_T\rangle
   & \equiv & \lim_{\beta_0 \rightarrow \infty} \frac{Z_{p}}{Z_{p,0}} \nonumber\\
   & =      & \sum_{C_n}\left(-\frac{U}{2}\right)^n \prod_{\sigma} \det M_{\sigma,p}(C_n)\;. \;\;\;\;
\end{eqnarray}
Here, the sum runs over all configurations $C_n$ as defined in Eq.~(\ref{Conf}) and 
Eq.~(\ref{Sum_Conf}). Note that $\beta$ in Eq.~(\ref{Sum_Conf}) has to be replaced by $
\Theta$. The matrices $ M_{\sigma,p} $ have precisely the same form as the matrices $
M_{\sigma} $, but since we have taken the limit $\beta_0 \rightarrow \infty $, the thermal
non-interacting Green's functions have to be replaced by the zero temperature ones:
\begin{equation}
G^0_{p,i,j} (\tau_1, \tau_2) = 
  \langle \Psi_T | T \hat{c}^{\dagger}_i(\tau_1) \hat{c}_j(\tau_2 ) | \Psi_T \rangle  \;.
\end{equation}
Hence, as in the Hirsch-Fye approach, the step from a finite temperature to
zero-temperature code is very easy and essentially amounts in replacing the finite
temperature non-interacting Green's functions by the zero temperature ones. However, there
is an important difference concerning measurements: measurements of observables which do
not commute with the Hamiltonian have to be carried out in the middle of the imaginary
time interval $[0,\Theta]$ to avoid boundary effects \cite{Assaad02}.
\begin{figure}[t]
\includegraphics[width=0.4\textwidth]{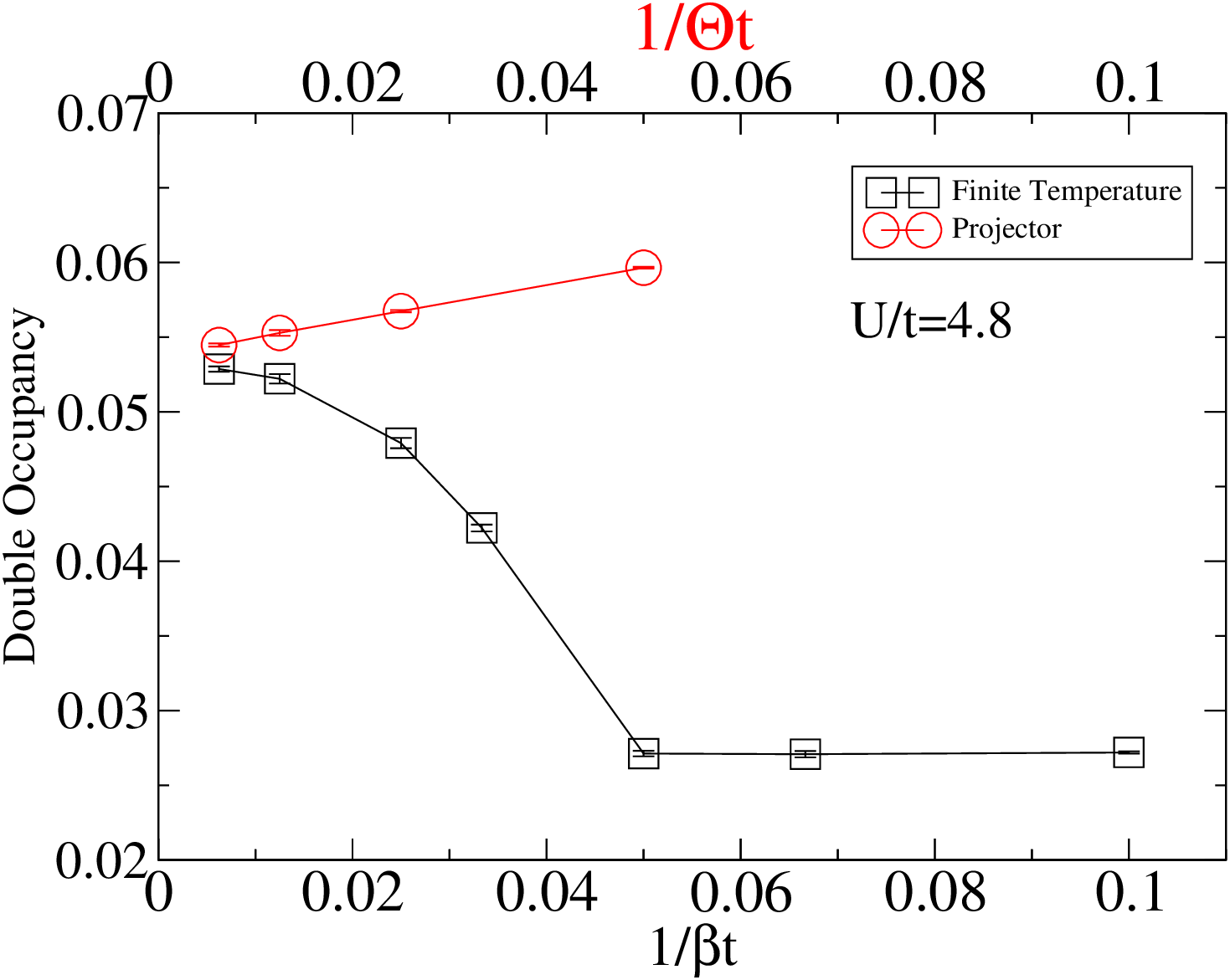} \\
\caption{ 
(Color online) DMFT calculation of the double occupancy for the half-filled Hubbard model 
as a function of i) temperature $1/\beta$ and ii) projection parameter $1/\Theta$. 
In the projective approach, observables are computed in the imaginary time range 
$[\Theta/4,3\Theta/4]$. This rather large measurement interval explains the slight 
discrepancy with the data of Ref. \cite{Feldbach04} for the projective code and at 
finite values of $\Theta$. However, extrapolation to $\Theta \rightarrow \infty $ where the 
measurement  range becomes irrelevant yields results consistent with Ref. \cite{Feldbach04}. }
\label{DMFT_T0.fig}
\end{figure}
We have tested this approach by reproducing Fig.~1 of the article \cite{Feldbach04} where
the projective Hirsch-Fye algorithm was incorporated in the DMFT self-consistency cycle. 
Fig.~\ref{DMFT_T0.fig} shows the results for the half-filled Hubbard model at $U/t = 4.8$,
density of states ${N(\omega)=\frac{8}{\pi W^2}\sqrt{W^2/4 -\omega^2}}$ and band-width
$W=4t$. Excellent agreement with the former Hirsch-Fye based results \cite{Feldbach04}
both at finite temperatures and in the limit ${\Theta \rightarrow \infty}$ were obtained. 
However, the diagrammatic approach allows to access much larger projection parameters
and/or lower temperatures. We refer the reader to Ref. \cite{Feldbach04} for the
implementation of the projective formalism in the self-consistency cycle.

\section{Application to the Hubbard-Holstein Model}

\label{Phonons_DDQMC}
Weak coupling DDQMC allows a very simple inclusion of phonon degrees of freedom. 
The path we follow here is to integrate out the phonons in favor of a retarded interaction, and 
then solve the purely electronic model with the DDQMC approach. Starting from the Hubbard-Holstein
model with Einstein phonons we show how to integrate out the phonons, describe some details of the 
algorithm and then present results within the DMFT approximation.

\subsection{Integrating out the Phonons}

The Hubbard-Holstein Hamiltonian we consider reads 
\begin{eqnarray}
  \hat{H} & = & -\sum_{ i,j,\sigma} t_{i,j} \hat{c}^{\dagger}_{i,\sigma}\hat{c}_{j,\sigma} + U \sum_{i}
                 \left(\hat{n}_{i,\uparrow}-1/2\right)\left(\hat{n}_{i,\downarrow}-1/2\right)\nonumber\\
          &   & + g \sum_i \hat{Q}_i \left(\hat{n}_{i} - 1\right)  
                + \sum_i \frac{\hat{P}_i^2}{2M} + \frac{k}{2} \hat{Q}_i^2 \;.
\end{eqnarray}
Here, $\hat{n}_i = \sum_{\sigma} \hat{n}_{i,\sigma}$ and the last two terms correspond 
respectively to
the electron-phonon coupling, $g$,  and the phonon-energy. The Hamiltonian
is written such  that for a particle-hole symmetric band, half-filling corresponds to
chemical potential $\mu = 0$. Opting for fermion coherent states
\begin{equation}
	\hat{c}_{i,\sigma}  | c \rangle = c_{i,\sigma} | c \rangle \;,
\end{equation}
 $c_{i,\sigma} $ being a Grassmann variable, and a real space representation for the
phonon coordinates
\begin{equation}
	\hat{Q}_i | q \rangle = q_i | q \rangle\;,
\end{equation}
the path integral formulation of the partition function 
reads
\begin{equation}	
	Z = \int \left[ {\rm d} q \right]  \left[ d c^{\dagger} d c\right] e^{- ( S_U + S_{\rm ep})} \;,
\end{equation}
with
\begin{eqnarray}
   S_U & = & \int_{0}^{\beta}  {\rm d} \tau    \sum_{i,j,\sigma} c^{\dagger}_{i,\sigma}(\tau) 
  \left( \delta_{i,j}\frac{\partial}{\partial \tau} - t_{i,j} \right)  c_{j,\sigma}(\tau) \nonumber\\
	 & & + \,U \sum_i ( n_{i,\uparrow}(\tau) -1/2) ( n_{i,\downarrow}(\tau) -1/2) \nonumber\\
   S_{\rm ep} & = &  \int_{0}^{\beta}  {\rm d} \tau 
   \sum_i \frac{ M \dot{q_i}^2(\tau)}{2} + \frac{k}{2} q_i^2(\tau) \nonumber\\
    & &  + \,g\, q_i(\tau) ( n_i(\tau) -1)\;. 
\end{eqnarray}
In Fourier space,
\begin{equation}
	q_{j}(\tau) = \frac{1}{\sqrt{\beta N} } \sum_{k,\Omega_m} 
   e^{-i ( \Omega_m \tau  -  k j ) } q_{k,m}\;,
\end{equation}
where $ \Omega_m $ is a bosonic Matsubara frequency, the electron phonon part of the
action reads
\begin{eqnarray}
& & S_{\rm ep} = \sum_{\Omega_m, k}  \frac{M}{2} \left( \Omega_m^2 + \omega_0^2 \right) 
       q^{\dagger}_{k,m} q_{k,m} + g q_{k,m} \rho^{\dagger}_{k,m}\;,  \nonumber \\
& &  \rho^{\dagger}_{k,m} = \frac{1}{\sqrt{\beta N}} \int {\rm d} \tau \sum_j 
e^{-i(\Omega_m \tau - k j)} ( n_{j}(\tau) - 1 )\,.
\end{eqnarray}
Gaussian integration over the phonon degrees of freedom leads to a retarded
density-density interaction:
\begin{eqnarray}
  \int \left[ {\rm d} q \right]  e^{-  S_{\rm ep} } = \quad\qquad\qquad\qquad\qquad\qquad\qquad\qquad\nonumber\\
     e^{\int_{0}^{\beta} {\rm d}\tau \int_{0}^{\beta} {\rm d}\tau'
               \sum_{i,j} \left[n_i(\tau )-1\right] D^0(i-j,\tau-\tau')
                          \left[n_j(\tau')-1\right]}\;. 
\end{eqnarray}
For Einstein phonons the phonon propagator is diagonal in real space,
\begin{eqnarray}
  & & D^0(i-j,\tau - \tau') = \delta_{i,j}\frac{g^2}{2 k} P( \tau - \tau')  \; \; {\rm with } 
   \nonumber \\ 
  & & P( \tau )  = \frac { \omega_0 } { 2 \left( 1-e^{-\beta \omega_0} \right) }
\left( e^{-|\tau|\omega_0} + e^{ - ( \beta - | \tau| ) \omega_0 } \right).
\end{eqnarray}
Hence the partition function of the Hubbard-Holstein model takes the form.
\begin{eqnarray}
       Z = \int \left[ d c^{\dagger} d c\right]\times \qquad\qquad\qquad\qquad\qquad\qquad\qquad\qquad\\
    \qquad e^{- (S_U  - \int_{0}^{\beta} {\rm d} \tau \int_{0}^{\beta} {\rm d} \tau' 
   \sum_{i,j} \left[ n_i(\tau) - 1\right] D^0(i-j,\tau - \tau')  \left[n_j(\tau') - 1\right])} \;.\nonumber
\end{eqnarray}
In the anti-adiabatic limit, $\lim_{\omega_0 \rightarrow \infty} P(\tau) =
\delta(\tau)$ such that the phonon interaction maps onto an attractive Hubbard
interaction of magnitude $g^2/k$. We are now in a position to apply the DDQMC algorithm
by expanding in both the retarded and Hubbard interactions.

\begin{figure}[t]
\includegraphics[width=0.4\textwidth]{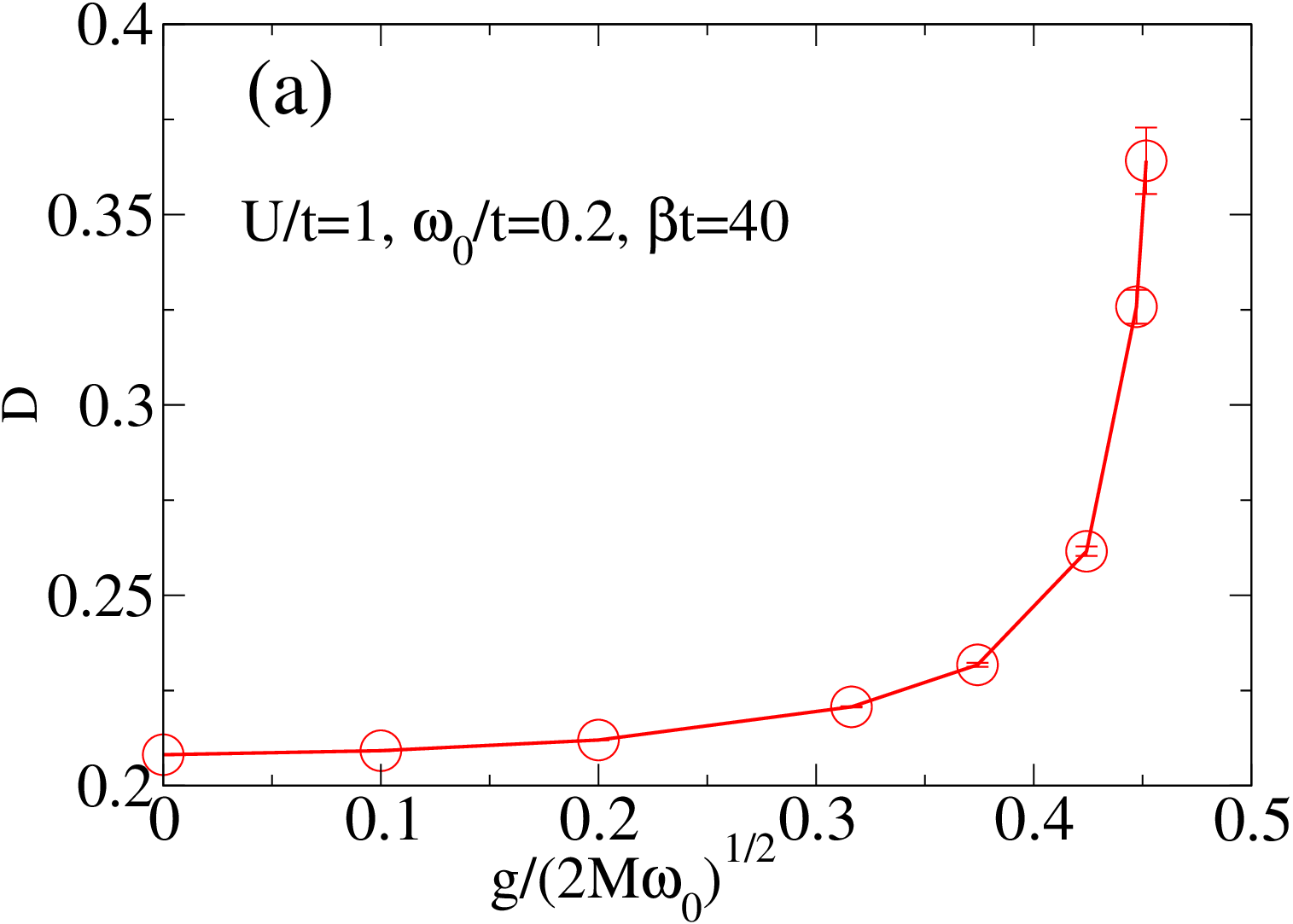} \\
\vspace{0.8cm} 
\includegraphics[width=0.4\textwidth]{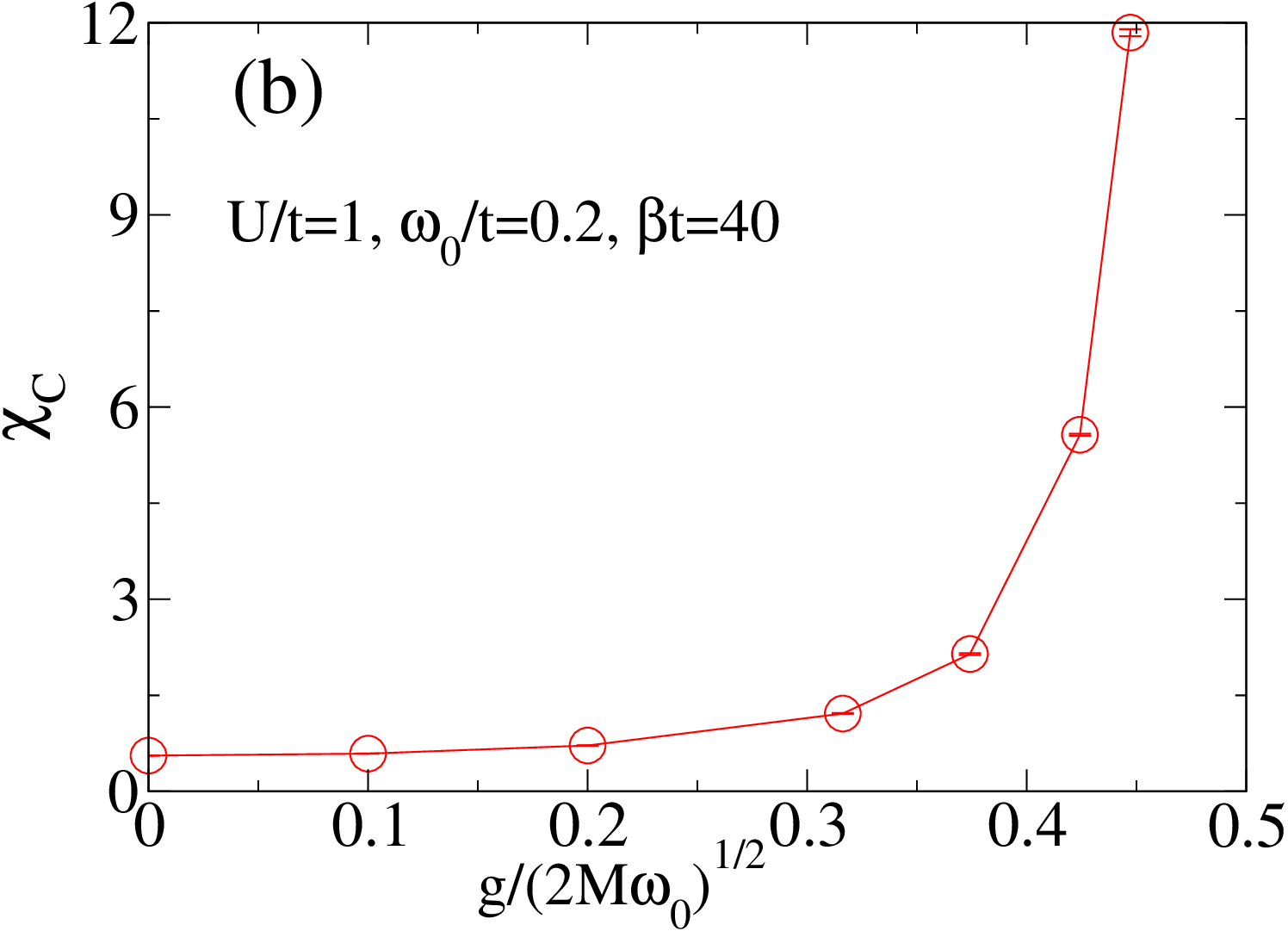} \\
\vspace{0.9cm} 
\includegraphics[width=0.4\textwidth]{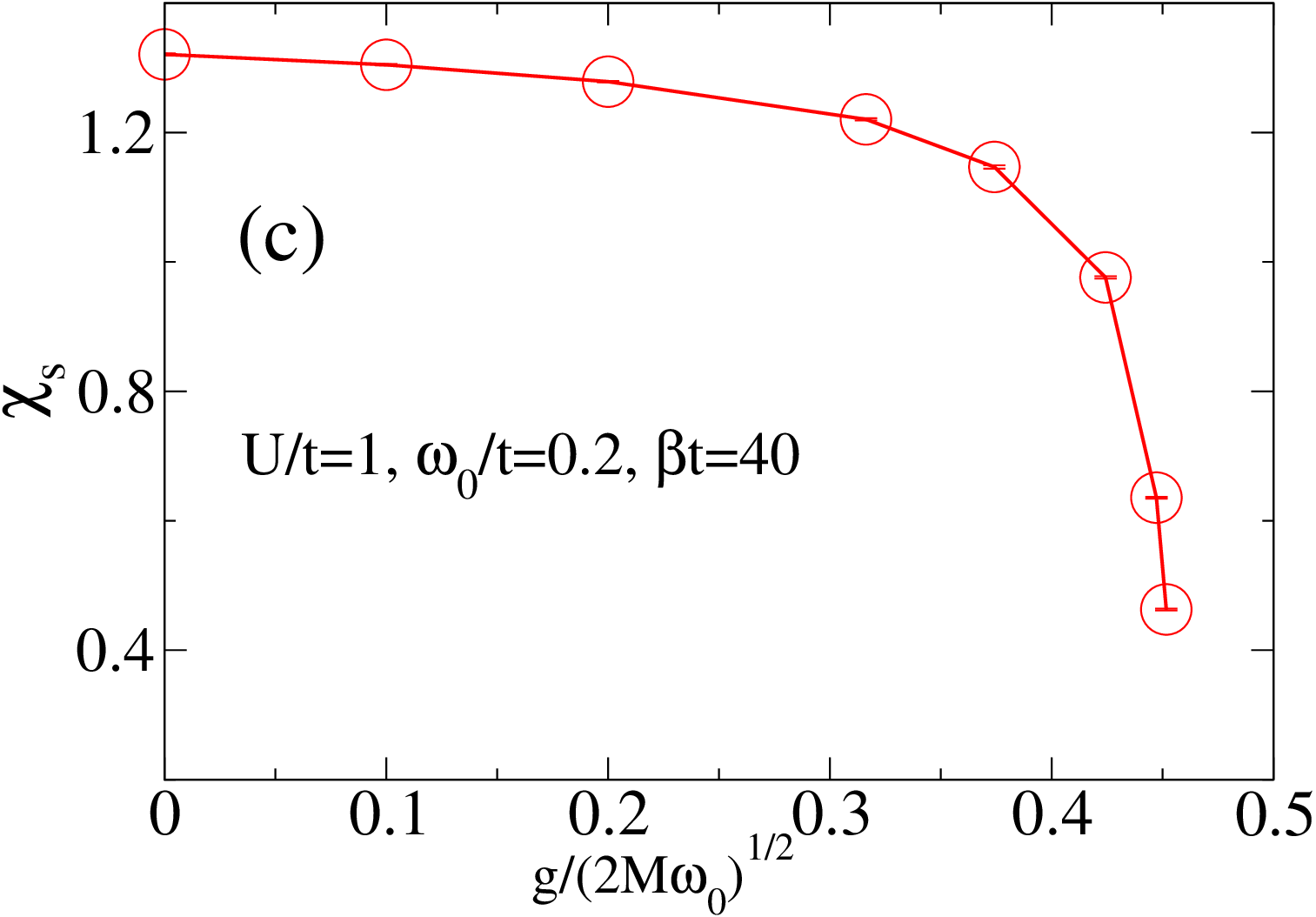} 
\caption{(a) Double occupancy, (b) local charge susceptibility and (c) local spin susceptibility
         for the Hubbard-Holstein model in the DMFT approximation. }. 
\label{DMFT_HH.fig}
\end{figure}

\subsection{Formulation of DDQMC for the Hubbard-Holstein model} 

To avoid the minus-sign problem at least for the one-dimensional chains we rewrite the
phonon retarded interaction as:
\begin{eqnarray}
   \label{Phonon}
   H_P(\tau) & & \!\!\!\!\!=  - \frac{g^2}{4k}\int_{0}^{\beta} {\rm d} \tau' \sum_{i,\sigma,\sigma'} 
         \sum_{s = \pm 1 } 
            P(\tau-\tau') \times \nonumber \\
 & &    
 \left[ n_{i,\sigma} (\tau ) - \alpha_{+}(s ) \right]   
 \left[n_{i,\sigma'} (\tau') - \alpha_{+}(s ) \right] . \nonumber \\
\end{eqnarray} 
For each phonon vertex, we have introduced an Ising variable: 
$s$. Summation over this Ising field reproduces, up to a
constant, the original interaction. Since the phonon term is attractive the adequate
choice of signs is ${\alpha_{+}(s) \equiv 1/2 + s \delta}$, irrespective of the spin
$\sigma$ and $\sigma'$. A similar argument as presented in Sec. \ref{Sign} then
guarantees the absence of a sign problem  for chains. Following Eq.~(\ref{Ising}) we
rewrite Hubbard term as
\begin{equation}
   H_U(\tau) = \frac{U}{2} \sum_{i,s } 
               \prod_{\sigma}( n_{i,\sigma}(\tau)  - \alpha_{\sigma}(s) )\;.
\end{equation} 
To proceed with a description of the implementation of the algorithm it is useful to
define a general  vertex
\begin{equation}
	V(\tau) = \left\{ i,\tau, \sigma, \tau', \sigma', s , b \right\} \;,
\end{equation}
where $b$ defines the type of vertex at hand, Hubbard (${b=0}$) or phonon (${b=1}$).
For this vertex we define a sum over the available phase phase 
\begin{equation}
	 \sum_{V(\tau)} =  \sum_{i,\sigma,\sigma',s,b} \int_{0}^{\beta} {\rm d} \tau'
\end{equation}
a {\it weight }
\begin{equation}
	w\left[V(\tau)\right]  =  \delta_{b,0} \frac{U}{2}  - 
                         \delta_{b,1} P(\tau-\tau') \frac{g^{2}}{4k}\;,
\end{equation}
as well as 
\begin{eqnarray}
  H & &\!\!\!\!\!\!\!\left[ V(\tau ) \right]  =   
	\delta_{b,0} \delta_{\sigma,\uparrow} \delta_{\sigma',\downarrow} \delta(\tau-\tau') 
          \nonumber \\
       & &   \left[ n_{i,\uparrow}(\tau)-\alpha_{+} (s ) \right] 
         \left[ n_{i,\downarrow}(\tau)-\alpha_{-} (s ) \right] +  \\
& &  \delta_{b,1} \left[ n_{i,\sigma}(\tau)  - \alpha_{+}(s) 
           \right]   
     \left[n_{i,\sigma'} (\tau') - \alpha_{+}(s) \right]. 
\nonumber 
\end{eqnarray}
With the above definitions, the partition function can now be written as
\begin{eqnarray}
	& & \frac{Z}{Z_0} = \sum_{n=0}^{\infty} (-1)^{n} 
\int_{0}^{\beta} {\rm d} \tau_1 \sum_{V_1(\tau_1)} w[V_1(\tau_1)] \cdots 
\int_{0}^{\tau_{n-1}} {\rm d} \tau_{n}   \nonumber \\
& & \sum_{V_{n}(\tau_{n})} w\left[V_n(\tau_{n})\right] \langle T \hat{H}\left[ V_1(\tau_1)\right] \cdots \hat{H} \left[ V_n(\tau_n)\right] \rangle_{0}\;.
\end{eqnarray}
As for the Hubbard model, a configuration consists of a set vertices ${C_n = \left\{
V_1(\tau_1), \ldots, V_n(\tau_n) \right\}}$. For a given configuration the  thermal
expectation value maps onto the product of two determinants in the spin up and spin 
down sectors.  The Monte Carlo sampling follows precisely the scheme  presented in Sec.
\ref{Monte-Carlo}, namely the addition and removal of vertices.

\subsection{Application to the Hubbard-Holstein model in the dynamical mean field approximation}

We have applied the above algorithm to the Hubbard-Holstein model within the  dynamical
mean-field theory approximation. We use a semicircular density of states, ${N(\omega) =
\frac{8} {\pi W^2} \sqrt{ W^2/4 -\omega^2}}$ with band-width ${W=4t}$.  Throughout this
section, we set ${U/t = 1}$, ${\omega_0 = 0.2 t}$, ${\mu=0}$, and use the finite
temperature implementation of the algorithm at ${\beta t = 40}$.  The choice  ${\mu =
0}$ corresponds to half-filling.  Fig.~\ref{DMFT_HH.fig}(a) plots the double occupancy, 
${D = \langle n_{i,\uparrow} n_{i,\downarrow} \rangle}$ as a function of the
electron-phonon  coupling.  To compare at best with the results of Ref. \cite{Koller04}
we write the phonon  coordinates in terms of bosonic operators, ${\hat{Q} = \frac{1} {
\sqrt{2M\omega_0} } \left( \hat{a} + \hat{a}^\dagger \right)}$, and plot our  results
as a function of
\begin{eqnarray}
	\tilde{g} = g/\sqrt{2M\omega_0} \;.  \\ \nonumber
\end{eqnarray}
Comparison of the results of Fig.~\ref{DMFT_HH.fig}(a) with those of
Ref.~\cite{Koller04}, show excellent agreement and a critical electron-phonon coupling
for the transition to the bipolaronic insulator at  ${\tilde{g}_c \simeq 0.45t}$. 

\begin{figure}[t]
   \includegraphics[width=0.4\textwidth,clip=true]{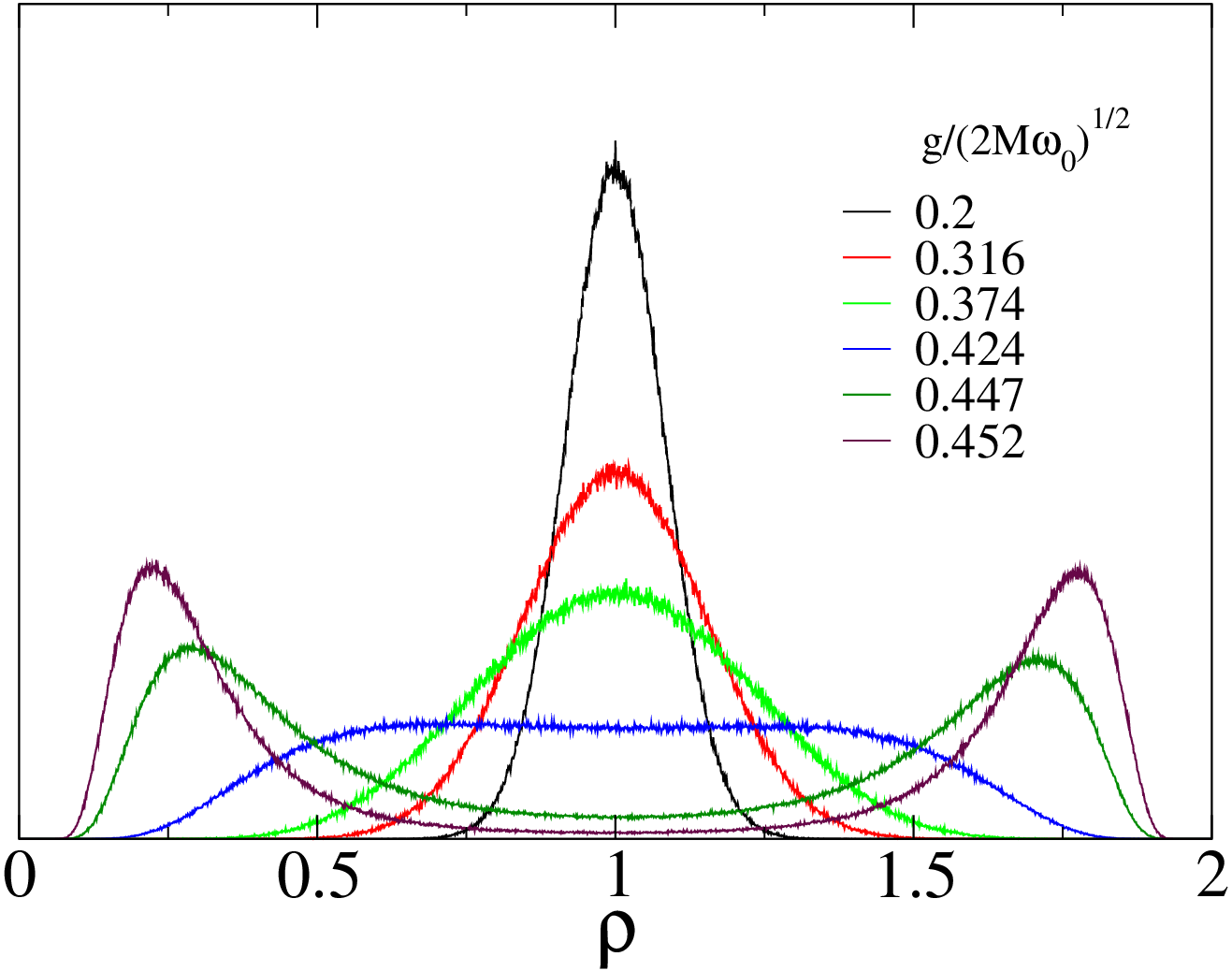}
   \caption{(Color online) The normalized histogram of the local density $\rho$ of the
         Hubbard-Holstein model at half-filling with ${U/t=1}$, ${\omega_0 = 0.2 t}$ and ${\beta t = 40}$, for
         several values of the electron-phonon coupling $g$.} 
   \label{Histogram_HH.fig}
\end{figure}

We have equally computed the local spin and charge susceptibilities:
\begin{equation}
	\chi_{\stackrel{C}{S}} = \int_{0}^{\beta} {\rm d} \tau \langle
	 \left[ \hat{n}_{i,\uparrow}(\tau) \pm \hat{n}_{i,\downarrow}(\tau) \right]
        \left[ \hat{n}_{i,\uparrow}  \pm \hat{n}_{i,\downarrow} \right] 
       \rangle \;.
\end{equation}
As can be seen in Fig.~\ref{DMFT_HH.fig}(b) in the vicinity the transition local charge fluctuations 
grow substantially and local spin fluctuations (see Fig. \ref{DMFT_HH.fig}(c)) are suppressed.  
The suppression of $\chi_S$ signals singlet pairing of polarons. The chemical potential 
$\mu = 0$ sets the average particle number $\rho = 1$ but the particle number itself oscillates 
strongly between an empty or doubly occupied site. This can be seen by taking histograms
as shown in Fig. \ref{Histogram_HH.fig}. In the vicinity of the transition a two peak structure
corresponding to a doubly occupied or empty site emerges. Close to the 
transition it becomes increasingly hard to guarantee a symmetric histogram -- corresponding to 
the particle-hole symmetry of the model -- and the simulation ultimately freezes 
in the doubly occupied or empty state.

\section{Conclusions}
We have presented two extensions of the diagrammatic determinantal method: projective 
schemes as well as the inclusion of phonons.    
In both cases we have tested successfully the 
approach for the Hubbard  as well as for the Hubbard-Holstein models 
in the dynamical mean-field approximation.  
The inclusion of phonons is not limited to Einstein modes and in principle any  dispersion relation 
can be easily implemented. 
One of the strong points of the weak-coupling DDQMC can be 
extended  very easily to larger clusters and used as a solver for cluster extensions of 
dynamical mean-field theories \cite{Maier05}. The crucial issue here is the severity  of the 
sign problem as the cluster size increases.  This is an issue which would need to be answered on 
a case to case basis.

\begin{acknowledgments}
We would like to thank N. Kirchner,  M. Potthoff and M. Troyer for conversations. We 
acknowledge financial support from the DFG under the grant number AS120/4-2. 
\end{acknowledgments}

\bibliographystyle{./prsty}

\end{document}